# Numerical investigation of nanosecond pulsed plasma actuators for control of shock-wave/boundary-layer separation

Kiyoshi Kinefuchi,[1,a)] Andrey Y. Starikovskiy[2)] and Richard B. Miles[2)]

[1]Research and Development Directorate, Japan Aerospace Exploration Agency, Tsukuba, Ibaraki 305-8505, Japan

[2]Department of Mechanical and Aerospace Engineering, Princeton University, Princeton, New Jersey 08544, USA

## Abstract

This study numerically explores the flow physics associated with nanosecond pulsed plasma actuators that are designed to control shock-wave induced boundary-layer separation in a Mach 2.8 supersonic flow. By using two dielectric barrier surface discharge actuator configurations, parallel and canted with respect to the flow velocity vector, a previous experiment suggested that the actuator worked in two ways to influence the interaction: boundary layer heating and vorticity production. The heating effect was enhanced with the parallel electrode and made the boundary-layer separation stronger, while the canted electrode produced vorticity and suppressed the boundary-layer separation due to the momentum transfer from the core flow. Because the detailed physical processes are still unclear, in this paper a numerical investigation is undertaken with a large eddy simulation and an energy deposition model for the plasma actuation, in which the dielectric barrier discharge produced plasma is approximated as a high temperature region. The flow characteristics without the plasma actuation correspond to the experimental observation, indicating that the numerical method successfully resolves the shock-wave/boundary-layer interaction. With the plasma actuation, complete agreement between the experiment and calculation has not been obtained in the size of shock-wave/boundary-layer interaction region. Nevertheless, as with the experiment, the calculation successfully demonstrates definite difference between the parallel and canted electrodes: the parallel electrode causes excess heating and increases the strength of the interaction, while the canted electrode leads to a reduction of the interaction strength, with a corresponding thinning of the boundary layer due to the momentum transfer. The counter flow created by the canted actuator plays an important role in the vortex generation, transferring momentum to the boundary layer and, consequently, mitigating the shock induced boundary layer separation.

## I. INTRODUCTION

Shock-wave/boundary-layer interactions (SWBLI) have significant influences on supersonic vehicles because they cause aerodynamic issues such as increased drag, high heat loading, and adverse pressure gradients[1], hence, the physics of SWBLI has been widely investigated both experimentally and numerically[2,3]. In isolators of high-speed airbreathing engines for example, the adverse pressure gradient caused by SWBLI sometimes induces flow separation, which can lead to a reverse flow in the boundary layer, and a decrease in total pressure resulting in significant performance loss or unstart of these engines. Therefore, the control of SWBLI is highly desirable and would be a key to realize more efficient space access or supersonic flight for the future. The authors have conducted experiments to mitigate the SWBLI with nanosecond surface dielectric barrier

[a)] Electronic mail: kinefuchi.kiyoshi@jaxa.jp.

discharge, NS-SDBD, plasma actuators[4]. The experimental result indicated that the NS-SDBD plasma actuator worked in two ways for SWBLI control: boundary layer heating and vorticity generation near the surface. The boundary layer heating made the SWBLI stronger and the size of the separation bubble increased. The vorticity generation appeared to successfully suppress the boundary-layer separation due to momentum transfer from the core flow to the boundary layer. The magnitude of the two effects of the NS-SDBD plasma has a strong relationship with the electrode configuration. The parallel electrode configuration in which the electrode was parallel to the flow leads to a strong heating effect. On the other hand, the SWBLI was suppressed due to momentum transfer from core flow to the boundary layer in the configuration in which the electrode was canted with respect to the flow.

The detailed mechanisms of vortex generation by NS-SDBD actuator and consequent SWBLI mitigation with the momentum transfer need to be better understood in order to improve the actuator performance for SWBLI control. Because experimental approaches to observe such physical process around the SWBLI region are complicated, time consuming and have diagnostics limitations, the development of a numerical approach is necessary to provide guidance to the development of effective methods for shock-wave/boundary-layer interaction control using dielectric barrier discharges. The characteristics of NS-SDBD actuators for both quiescent flow and subsonic flow has been experimentally and numerically investigated in the following literature: Starikovskii[5] showed that an overheating occurs in the NS-SDBD region due to fast (< 1 μs) thermalization, and that the shock wave together with the secondary vortex flows disturbs the main flow, which causes a momentum transfer into the boundary layer. Unfer[6] developed a numerical model coupling fluid discharge equations with compressible Navier-Stokes equations. The results suggested fast gas heating occurs in the boundary layer, consequently, a shock wave is produced, which agrees with the experiment conducted by Roupassov[7]. Some kinetic models have been proposed and demonstrated such physical process for a NS-SDBD actuator[8-11]. The full kinetic energy deposition model approach requires huge computational cost, hence, instead of that, simpler energy deposition models have been proposed to avoid the numerical complexity. In these energy deposition models, only the power density or temperature rise generated by the actuator are considered, eliminating the plasma kinetic modeling. Abdollahzadeh[12] demonstrated that an energy deposition model based on a plasma kinetic model is able to capture the main features of the effect of NS-SDBD actuators correctly with less computational time. Popov[13,14] numerically investigated NS-SDBD actuators for laminar flow on a flat plate based on an energy deposition model. Gaitonde[15-17] surveyed the spatial temperature rise distribution on an energy deposition model with a Gaussian profile and found that it reproduces the shock wave features observed in the experiment by Takashima[18]. Gaitonde demonstrated the effectiveness of Large Eddy Simulation (LES) coupled with the energy deposition model to resolve vorticity and the disturbances induced by the NS-SDBD actuator with realistic computational resources. As for applications of NS-SDBD actuators in the supersonic regime, the number of reports is limited. Yan[19] numerically studied pulsed thermal perturbation in Mach 1.5 supersonic flow. The perturbation produces streamwise vortical disturbance and induces laminar to turbulent transition.

This study explores the underlying physics associated with NS-SDBD actuator for control of SWBLI based on numerical analysis. The numerical setup is in accordance with the previous experiment by the authors[4]. Turbulence is treated as combined approach of RANS (Reynolds Averaged Navier-Stokes) and LES: LES is applied around the NS-SDBD region to resolve the vorticity generation and evolution, followed by momentum transfer to the boundary layer, while RANS is applied to the rest of calculation domain to save the computational resources. The NS-SDBD effect is described based on an energy deposition model. First, the numerical model without plasma actuation is discussed and validated by comparison with the experimental results of wall pressure distribution, schlieren images and velocimetry around separation region. Then, the effect of NS-SDBD actuation is investigated in detail in the parallel and canted electrodes configurations in terms of the momentum transfer and the heating effect. Finally, the mechanism of SWBLI mitigation by the canted electrode configuration is summarized.

## II. NUMERICAL METHOD

Figure 1 depicts a schematic of the computational domain with a coordinate system based on the experiment[4]. The blowdown supersonic wind tunnel that consists of an indraft supersonic nozzle and an ejector system was used in the experiment. A supersonic flow is obtained in the observation channel located downstream of the supersonic nozzle. The observation channel is 150 mm in length with rectangular cross section, 30.5 mm in height and 51 mm in width. A 14 degrees wedge shape shock generator with 6 mm in height is installed at the top of the observation channel. The computational domain in Fig. 1 was modeled based on the observation channel. The incident shock wave interacts with boundary layer on the bottom surface, and produces a reflected shock wave. The measured suction pressure of the observation channel before the interaction region was 27 torr, and the corresponding Mach number and mass flux are 2.8 and 68 kg/s/m$^2$, respectively. The boundary layer is turbulent as the Reynolds number is approximately order of $10^5$. Seven pressure taps locate at the spanwise center and the bottom of the observation channel to measure pressure distribution in the SWBLI region, and two quartz windows are located on the sidewalls for optical access to the SWBLI region. The pressure and optical data will be compared with the numerical result.

The geometries of two electrode configurations, parallel electrode and canted electrode are also described in Fig. 1. The electrodes used to generate the NS-SDBD plasma were installed on the bottom wall and upstream of the SWBLI region. The parallel electrode was installed along the core flow as long as possible expecting enhancement of the effect of plasma actuation.

The resulting discharge length was 80 mm. The canted electrode has the same discharge length and margins from both sides, 13.5 mm, to avoid the interaction between the side walls and discharge. As a result, the canted angle of the canted configuration was 18 degrees against the core flow direction. The actuator was driven by a nanosecond-pulse generator in the experiment and the pulse width is 30 ns.

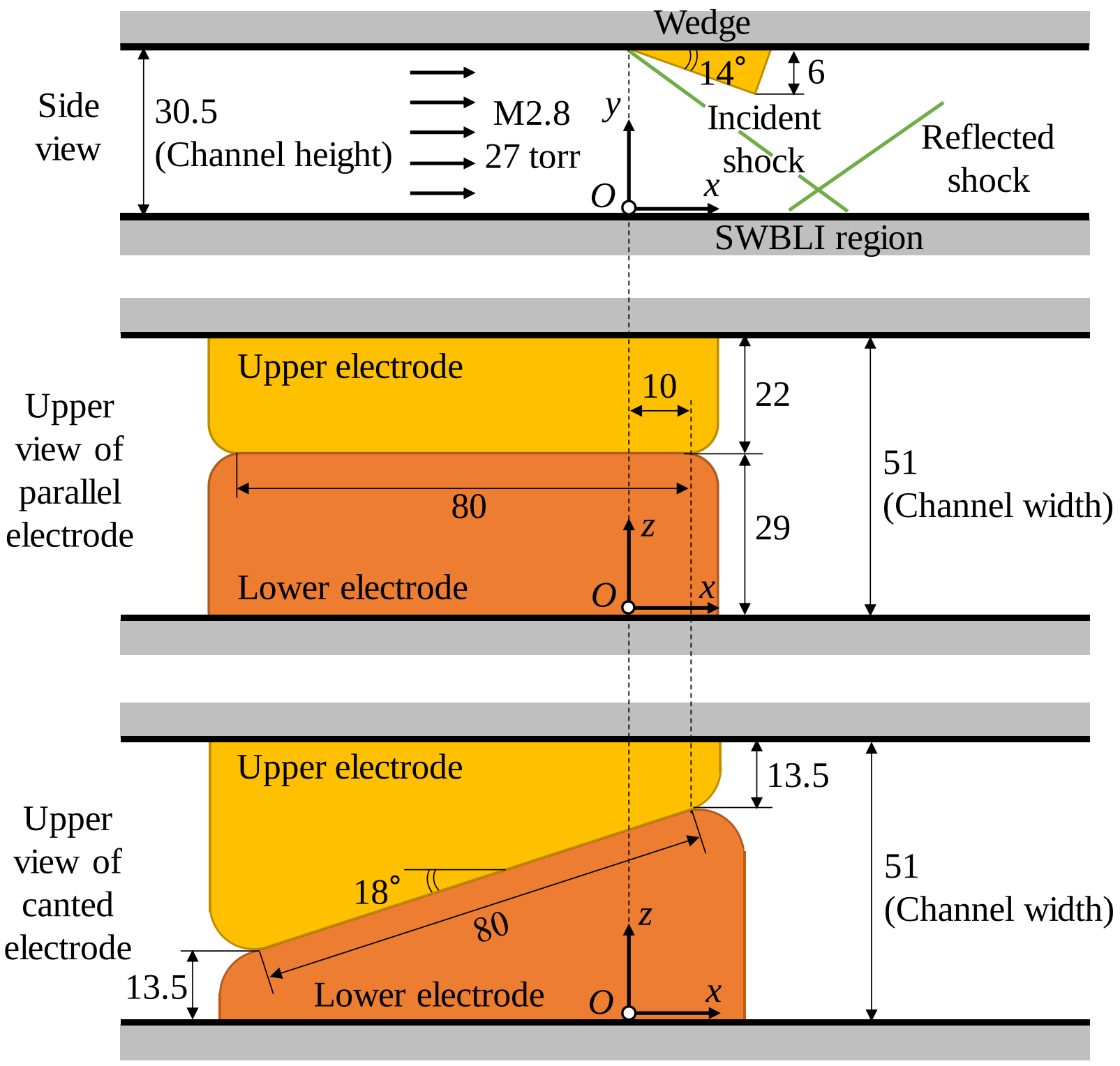

FIG. 1. Schematic of computational domain with coordinate system (dimensions are in mm).

Figure 2 describes the mesh distribution, with the numerical setup based on the schematic shown in Fig. 1. The structured hexahedral mesh consists of approximately 8,400,000 cells. The inlet boundary condition is specified as the mass flux, 68 kg/s/$m^2$ with total temperature of 290 K. According to the experimental observation[20], the boundary layer thickness just upstream of SWBLI region is 4 mm. The boundary layer thickness at the inlet boundary is set to 2 mm to be consistent with the experiment. No-slip condition is applied to all the walls. The governing equations consist of the unsteady compressible Navier-Stokes equations with a RANS and LES hybrid turbulence treatment as shown in Fig. 3: Around the NS-SDBD region, LES was applied with the finer mesh as seen in Fig. 2 to resolve vorticity followed by the pulsed plasma discharge. The mesh spacing in the *y*-direction (boundary layer thickness direction) on the bottom wall is 0.02 mm. The LES region on the bottom wall do not include the SWBLI region because the SWBLI region interacts with the side walls[21] and the LES could not resolve

SWBLI due to insufficient mesh spacing on the side walls. The objective of this study is focusing on the physics around the NS-DBD plasma actuator, hence, RANS with a $k$-$\varepsilon$ turbulent model is used in the rest of the computational domain including SWBLI region with a coarse mesh to solve the problem with realistic computational resources. The RANS approach has successfully captured flow characteristics around SWBLI region as shown later with the experimental results without the plasma actuation. The time step size for the calculation is set to 1 μs to resolve the flow characteristics including the shock structures in this mesh distribution.

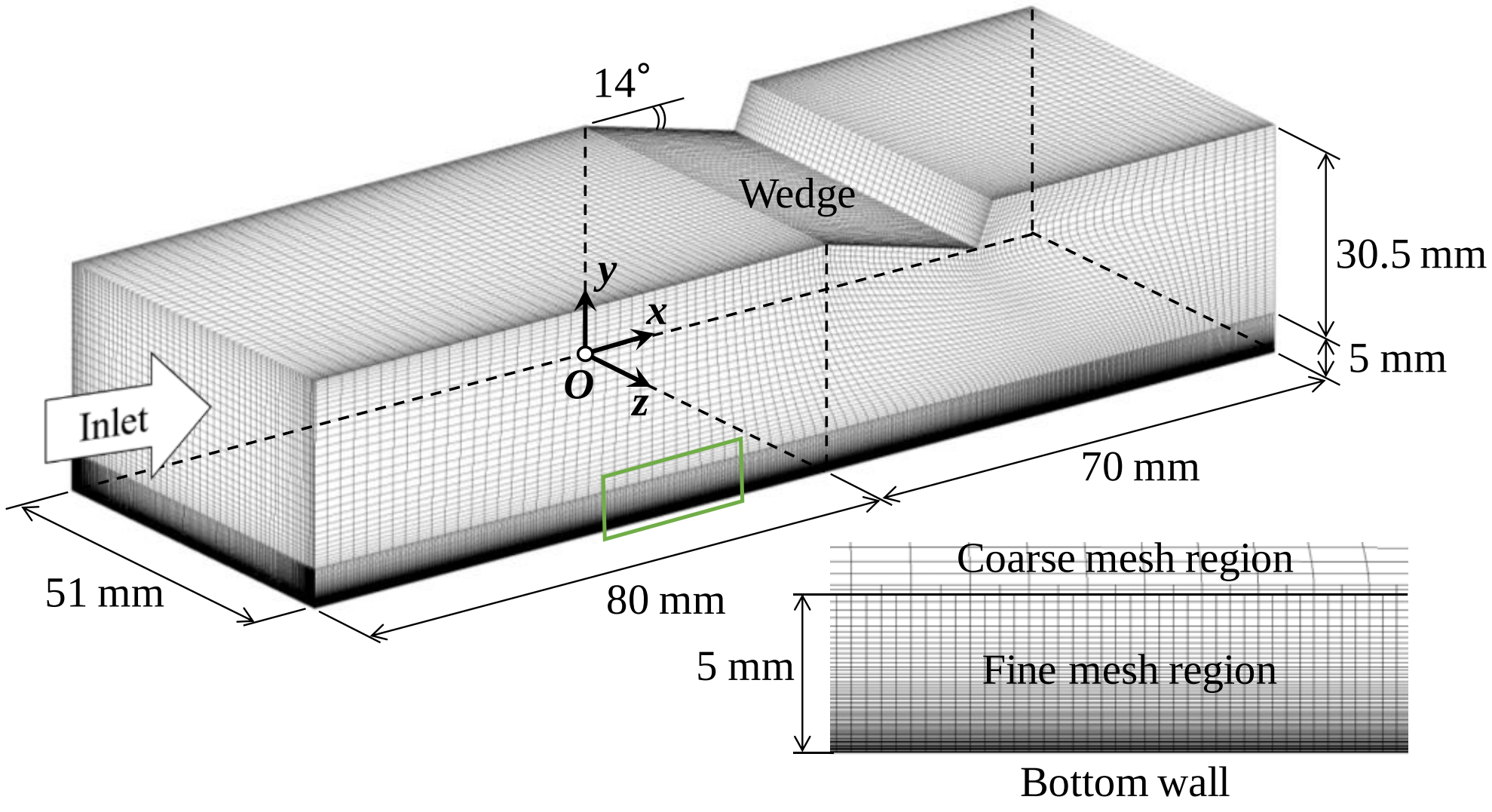

FIG. 2. Mesh distribution and numerical setup.

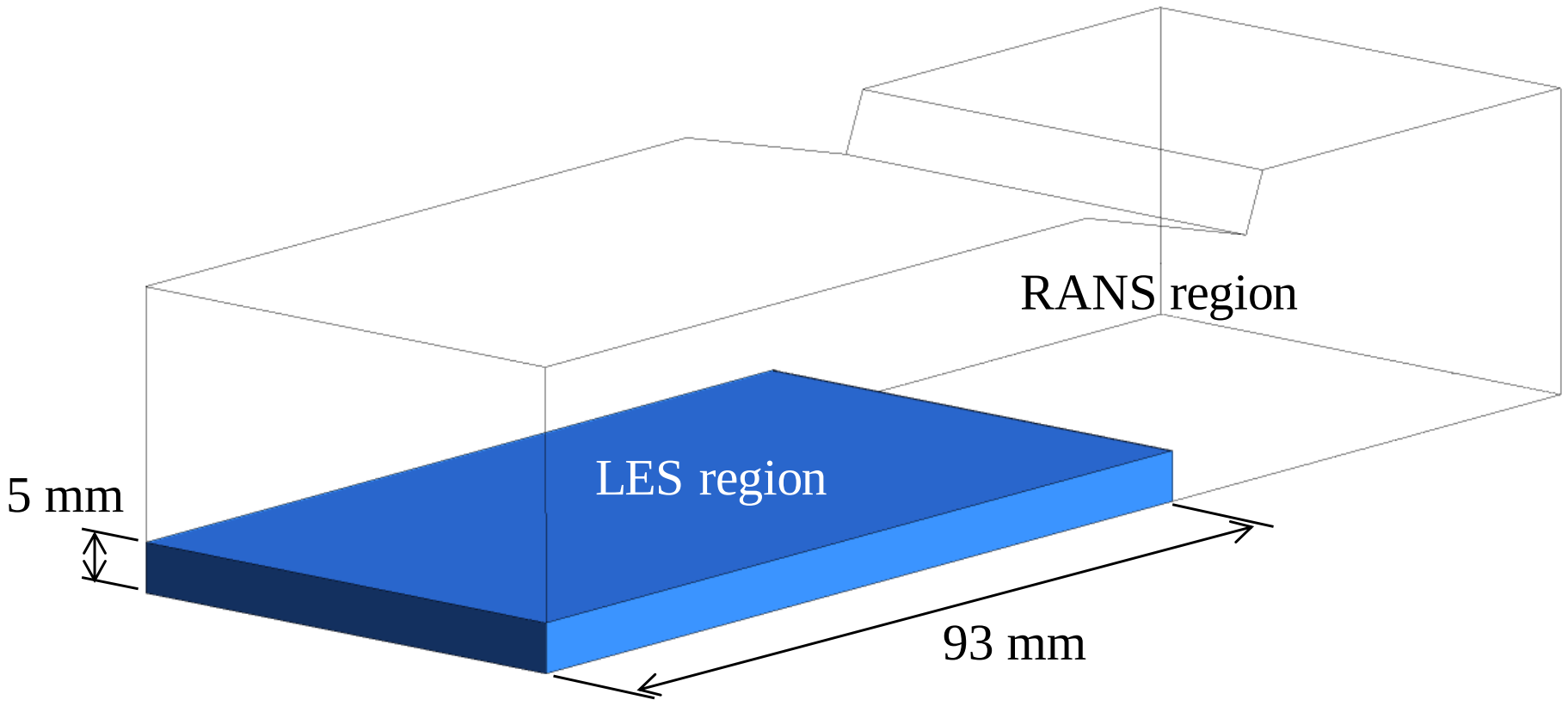

FIG. 3. Distribution of LES region (colored) and RANS region.

The NS-SDBD plasma actuator was numerically implemented based on an energy deposition model; hence, the temperature rise $\Delta T$ is imposed in the NS-SDBD plasma region. The relationship between $\Delta T$ and deposited energy $Q$ to the plasma is expressed as follows

$$Q = \int \rho C_v \Delta T dV \quad (1)$$

where $\rho$ is air density, $C_v$ is specific heat at constant volume of air, and $dV$ is volume element in the plasma region: The integral is carried out inside the plasma region. The deposited energy $Q$ is set to be 50 mJ in accordance with the experiment. The NS-SDBD actuators were operated in a low pressure environment, 27 torr in the experiment, thus, the plasma distribution was not the same as in the atmospheric condition. The observation of the NS-SDBD plasma under various surrounding pressure by Starikovskiy[22] showed the plasma length and height was inversely proportional to square root of surrounding pressure. Here, the shape of plasma region is simply assumed to be a 1/4 cylinder with 20 mm in radius. The radial distribution of $\Delta T$ in the 1/4 cylinder-shaped plasma is assumed to have Gaussian profile similar to the investigation of Gaitonde[17] as follows.

$$\Delta T / T_{\mathrm{ref}} = exp\left[-n\left(\frac{r}{R}\right)^2\right] \quad (2)$$

Where $r$ is radial coordinate and $R$ is the radius of the 1/4 cylinder, 20 mm. $T_{\mathrm{ref}}$ should be determined to satisfy Eq. (1). $n$ is constant and set to be 5 to achieve smooth distribution. The radial distribution of temperature rise is depicted in Fig. 4.

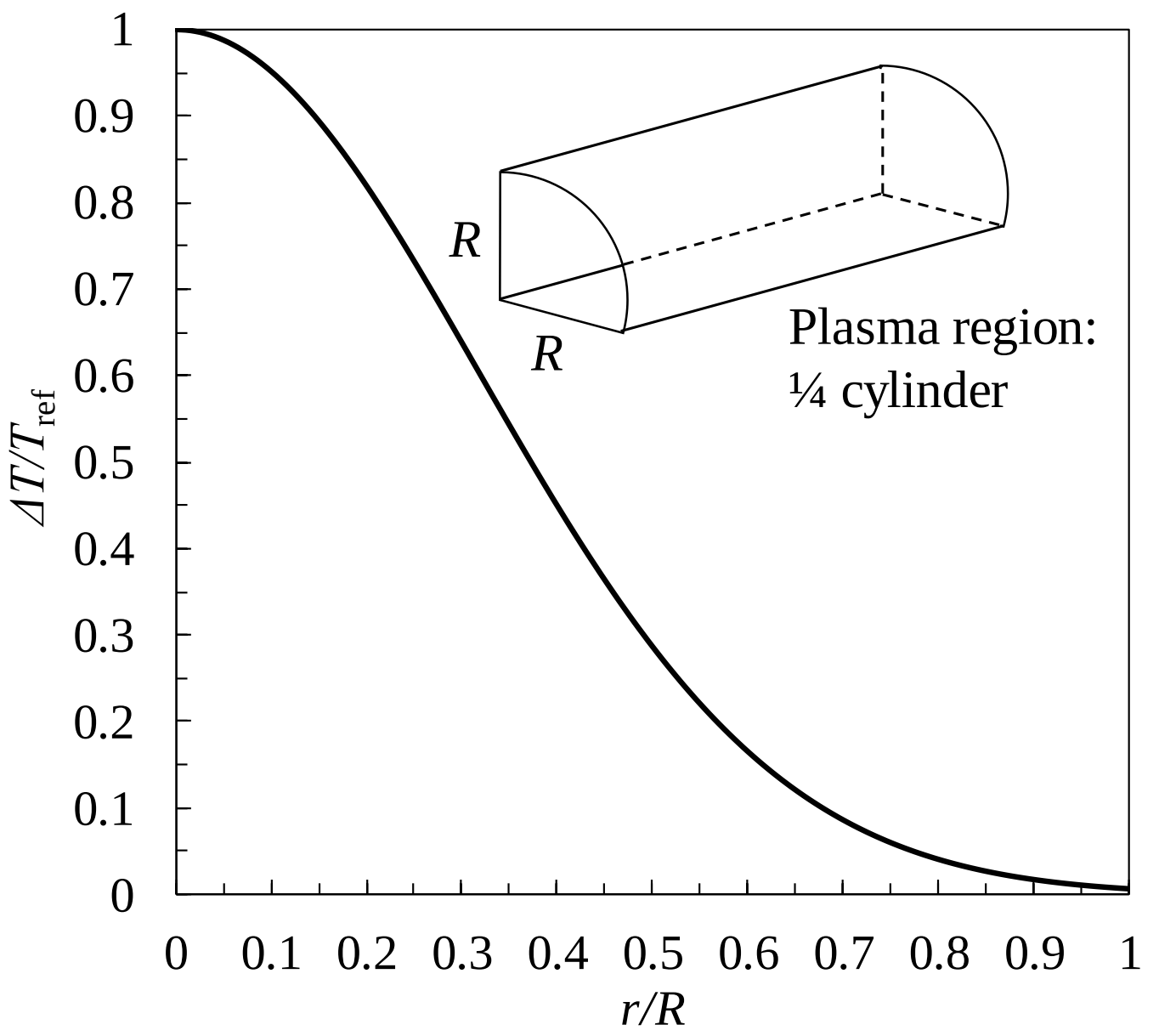

FIG. 4. Temperature rise distribution in NS-SDBD region.

The physical sequence of flow heating by NS-SDBD plasma is as follows[23]; 1) electrons gain some energy from the external electric field; 2) electrons excite/dissociate/ionize the molecules/atoms; 3) recombination and collisional quenching lead to energy release from internal to translational degrees of freedom. These processes are completed in less than 1 μs, much faster than flow response time; therefore, the evolution of thermalization is neglected and the following numerical procedure is adopted. First, a solution without plasma actuation is calculated and then the temperature rise based on Eqs. (1) and (2) is imposed onto the plasma region of the solution. Then, solutions with plasma actuation are obtained by using this as an initial condition. The pulse frequency of plasma actuation ranged from 100 Hz to 10 kHz in the experiment and the effect of the actuation to the flow longed in roughly 600 μs. That means each pulse is not overlapped at less than 1.7 kHz pulse frequency. The numerical calculation considers only one pulse, which corresponds to the cases without the overlap of pulses, or less than 1.7 kHz pulse frequency.

## III. RESULT AND DISCUSSION

### A. Flowfield without plasma actuation

To check the validity of the numerical method and investigate the flowfield around the SWBLI region, the numerical solution without plasma actuation was compared with the experimental results. Figure 5 shows $x$-velocity, Mach number, pressure, and temperature distributions at the spanwise center plane without the plasma actuation. The incident shock, reflected shock and the SWBLI region are observed.

Figure 6 describes pressure distribution on the bottom wall at the spanwise center. The measured data corresponds to the calculation result. When boundary-layer separation occurs, a constant pressure region called a pressure plateau is observed. A weak pressure plateau could be identified from the fact that the pressure gradient becomes smaller at around $x$ = 30 to 40 mm in Fig. 6. This indicates a weak separation regime. It should be noted in Fig. 6 that the numerical plot is not smooth at $x$ = 13 mm, which corresponds to the boundary between LES and RANS regions shown in Fig. 3. Hence, the unsmooth appears to be caused by numerical error.

The numerical contour of $d\rho/dx$ for a cross section at the spanwise center is shown in Fig. 7 with the schlieren image taken from the experiment. These also show good agreement: the incident shock, reflected shock and SWBLI region are clearly seen in both pictures. The rapid wall pressure rise before the pressure plateau in Fig. 6 induces compression waves in the supersonic part which turn into the reflected shock as seen in Fig. 7. The incident and reflected shock intersect each other, and the incident shock curves in the boundary layer due to the decrease in the local flow velocity. Finally, the incident shock disappears on the sonic line in the boundary layer and forms an expansion fan. The angle of incident shock is approximately 33 degrees in both cases and that corresponds to the estimated angle for Mach 2.8. It should be noted that the location of the pressure plateau in Fig. 6 corresponds to the SWBLI zone in these pictures.

Femtosecond laser electronic excitation tagging (FLEET)[24] velocimetry was applied to measure the velocity vector around the SWBLI region in the experiment. Figure 8 shows the numerically predicted velocity vector at the spanwise center around the SWBLI region along with the FLEET result. The calculation depicts a separation bubble, flow circulation including backflow in the SWBLI region, as expected from the existence of the pressure plateau. The FLEET measurement and numerical calculation show the similar results, but a slight difference in the vector directions because of the large spatial velocity gradient around SWBLI region.

In summary of the investigation of flow calculation without plasma actuation, the numerical approach with the RANS/LES hybrid method appropriately describes the flowfield, including the incident shock and reflected shock, and the SWBLI region as well as the separation bubble.

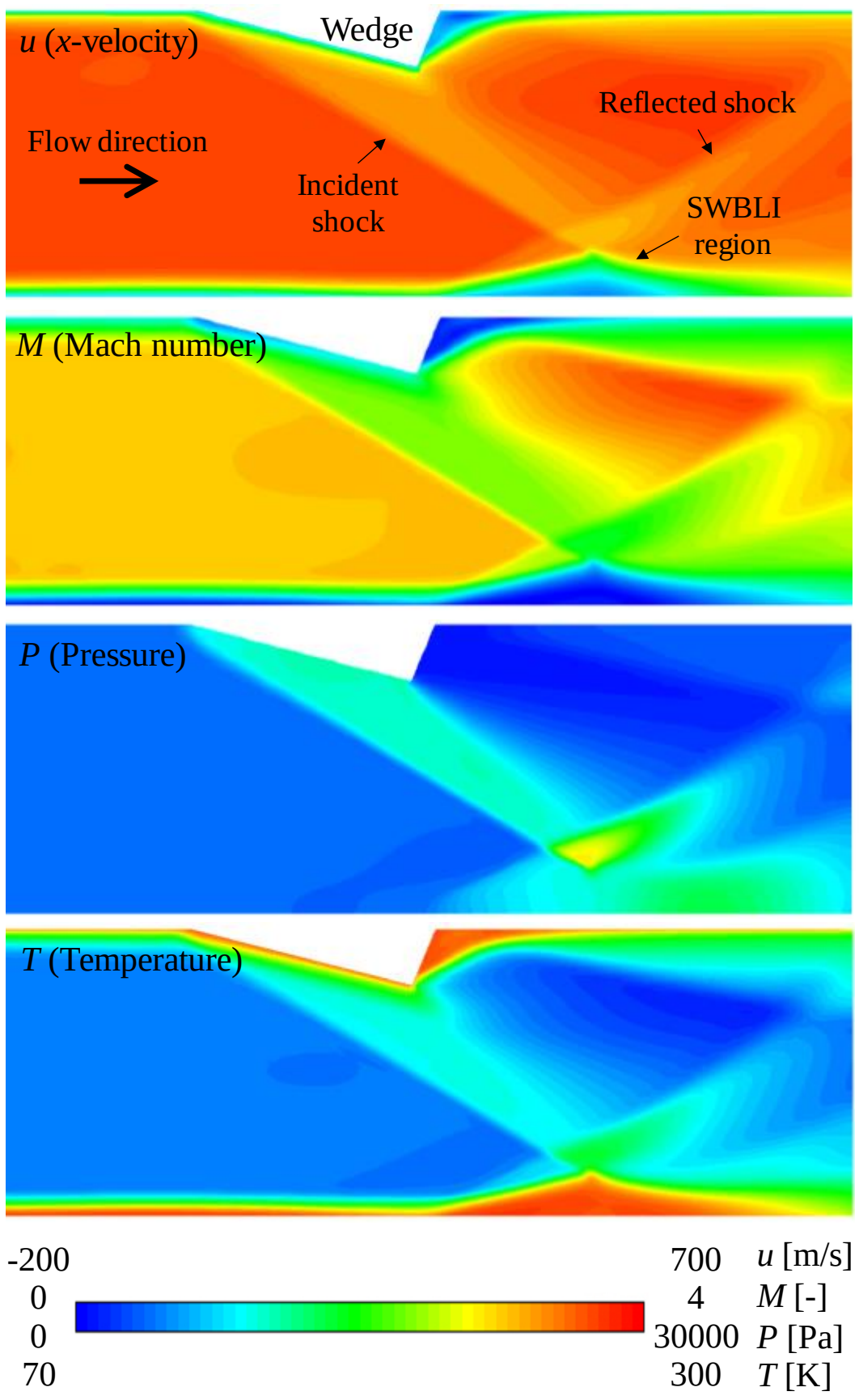

FIG. 5. *x*-direction velocity, Mach number, pressure, and temperature distribution at spanwise center without plasma actuation.

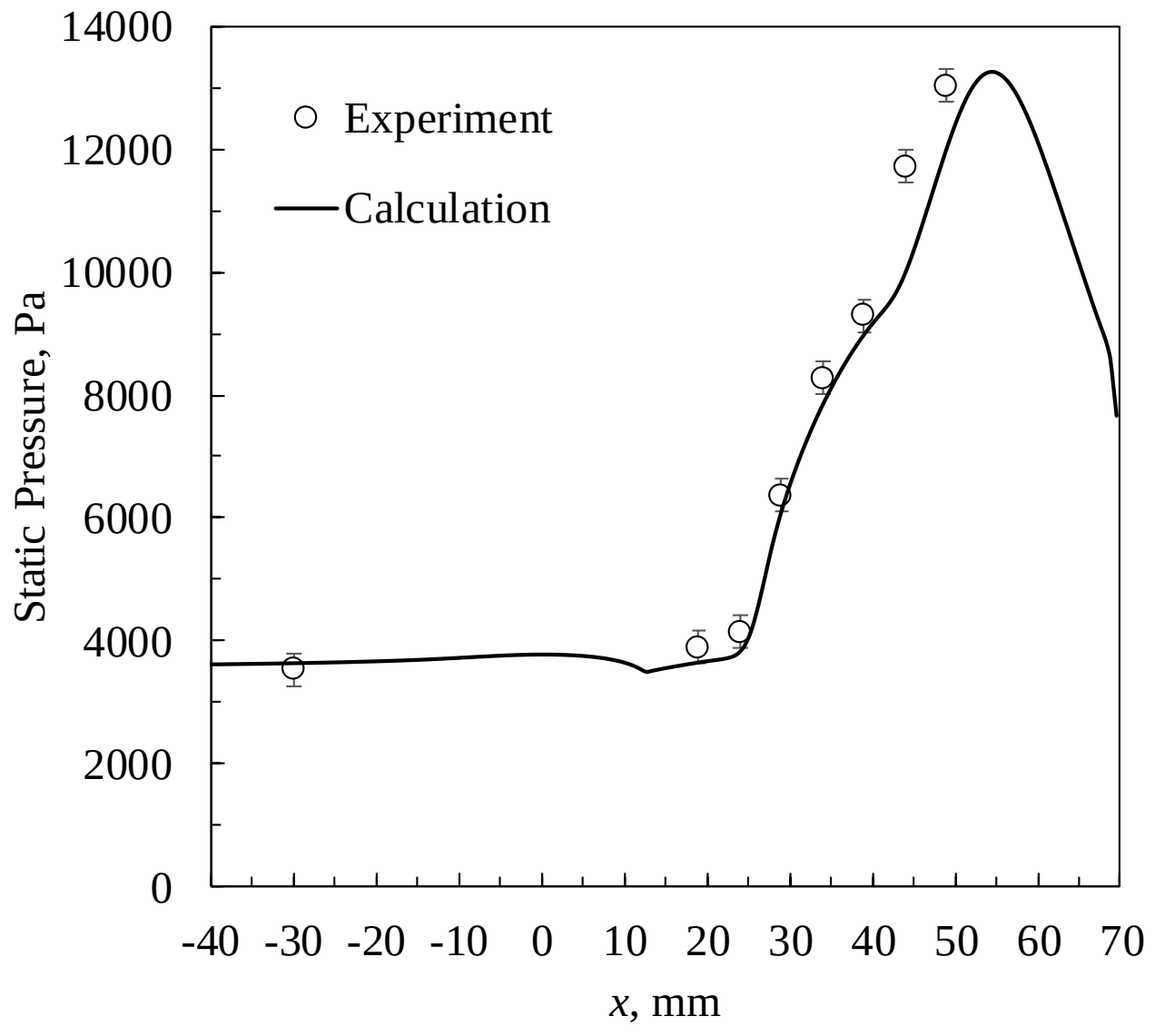

FIG. 6. Comparison of pressure distribution on bottom wall without plasma actuation at spanwise center ($z$ = 25.05 mm) between experiment and calculation

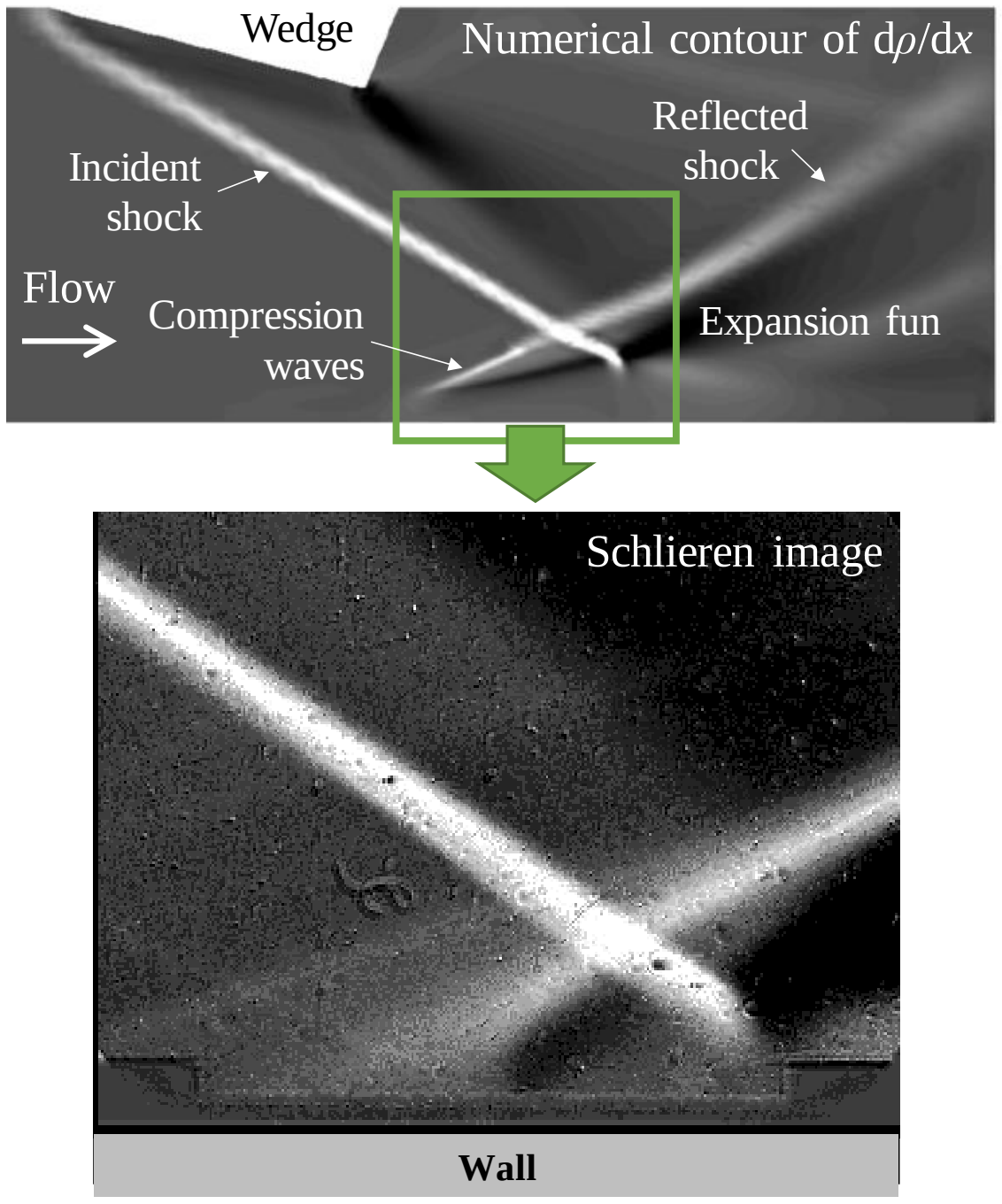

FIG. 7. Comparison of numerical contour of d$\rho$/d$x$ at cross section of spanwise center and schlieren image in experiment

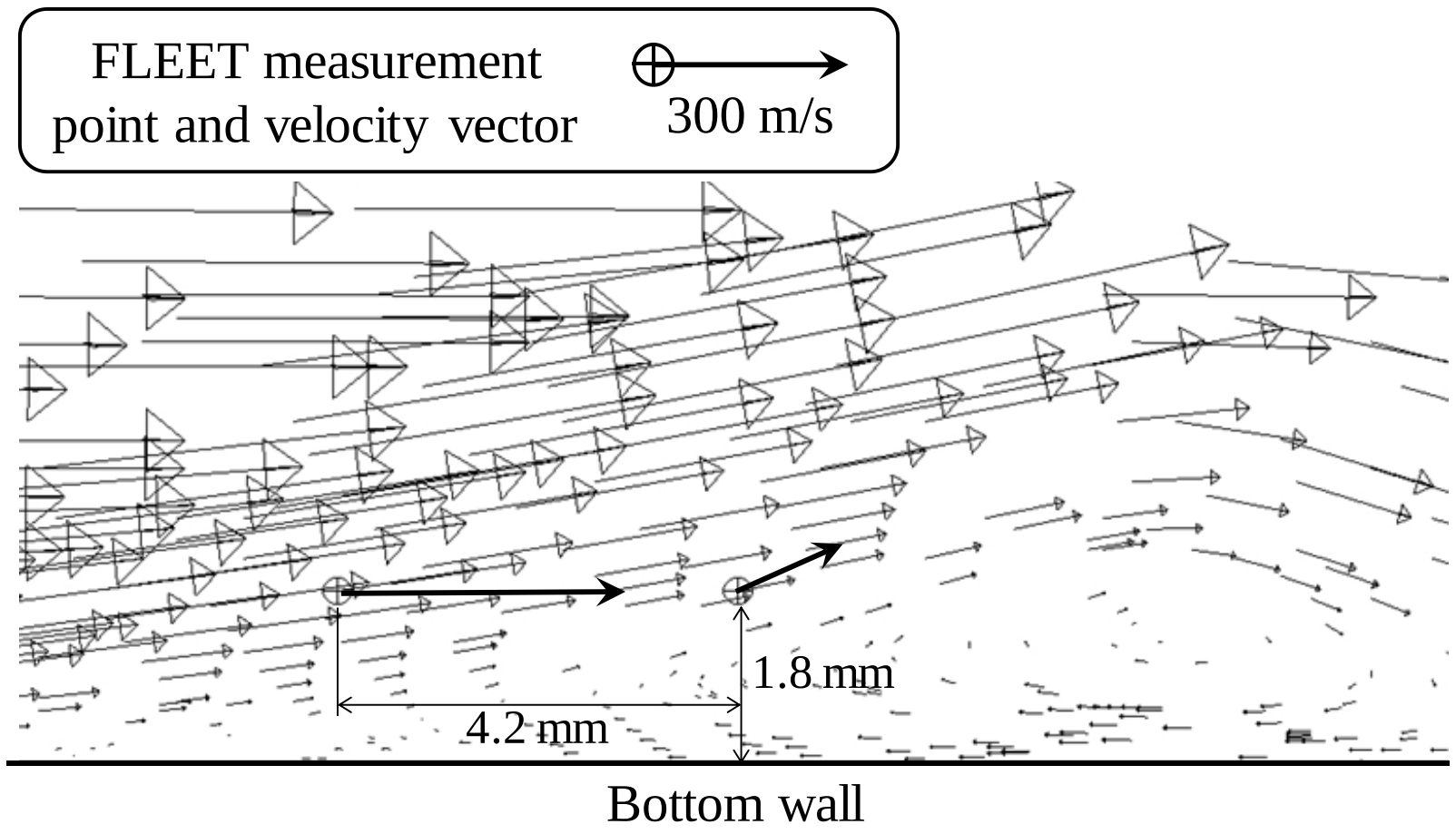

FIG. 8. Comparison of velocity vector around SWBLI region between calculation and measurement by FLEET.

## B. Investigation of plasma actuation

### 1. Characteristics of high temperature region

The initial temperature distribution that simulates the flow thermalization by the NS-SDBD plasma is shown in Fig. 9 in the parallel electrode case. That temperature distribution is parallel to the flow velocity vector for the parallel electrode case, and it is canted at 18 degrees in the canted electrode case, as seen in Fig. 1. Figure 10 describes pressure evolution from the high temperature region at $x = 0$ mm cross section in the parallel electrode case. At $t = 0$ µs, the pressure distribution corresponds to the initial temperature distribution shown in Fig. 9. Then, a shock wave followed by an expansion wave generated from the high temperature plasma region propagates perpendicular to the electrodes. The propagation speed of the shock wave is 220 m/s, which corresponds to the measured shock speed from the movie in the experiment[4]. Figure 11 shows an isometric view of the temperature evolution in the parallel electrode case. The high temperature region moves with the stream: in the boundary layer, the speed of the high temperature region is slow, while it moves faster in the core flow. The high temperature region lowers the Mach number, and that causes the SWBLI region to become larger with the high temperature region flowing into it. The shock wave propagation and the high temperature region movement have similar characteristics as well as in the canted electrode case, except at the canted angle of 18 degrees with respect to the main flow.

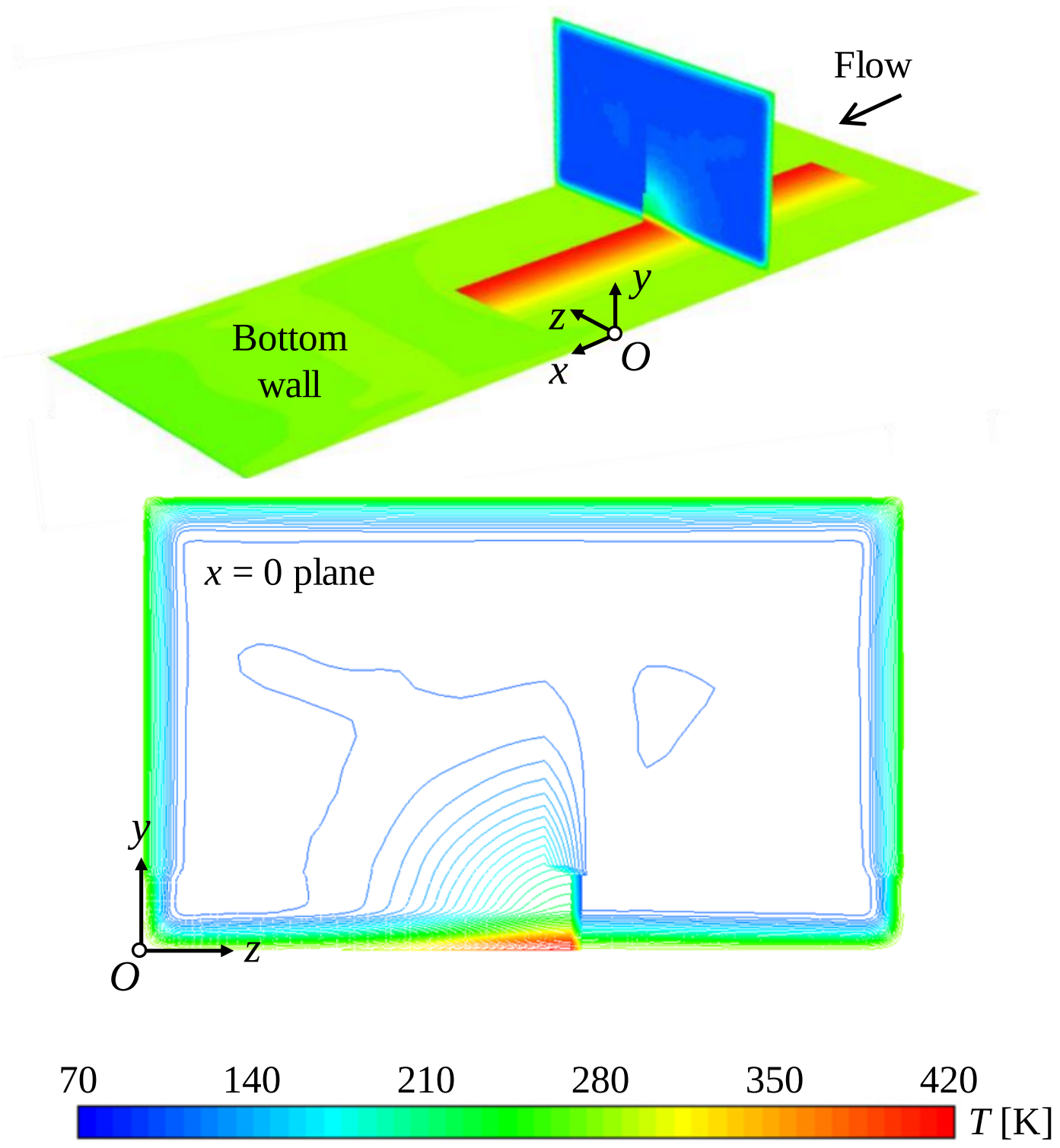

FIG. 9. Initial temperature distribution in parallel electrode case.

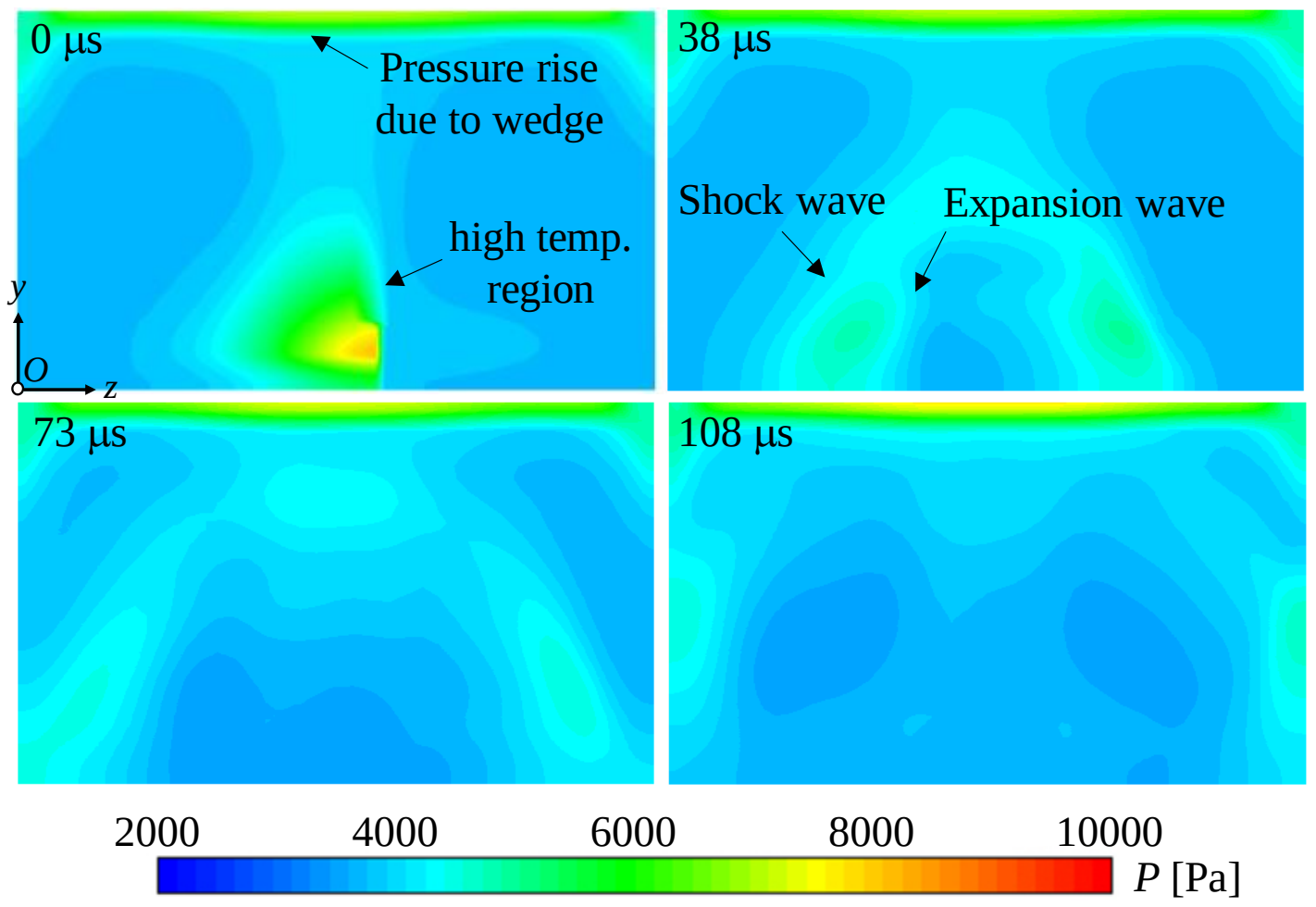

FIG. 10. Pressure evolution at $x$ = 0 mm cross section in parallel electrode case.

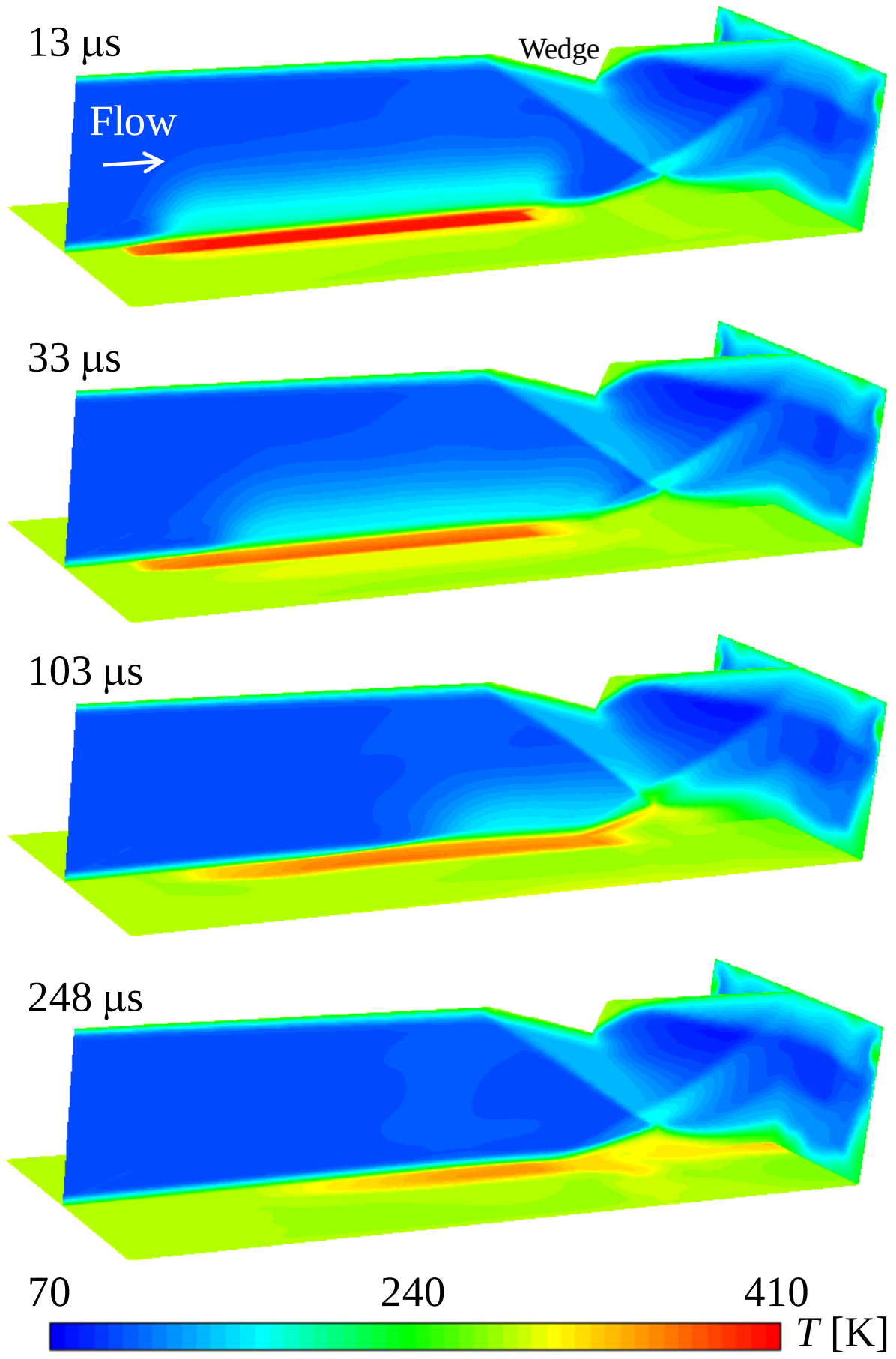

FIG. 11. Isometric view of temperature evolution in parallel electrode case.

### *2. Actuation effect to interaction region and boundary layer*

Figure 12 shows comparison of the reflected shock movement, indicating the size of SWBLI region for both the calculation and the experiment. The experimental plot corresponds to the data with pulse frequency of 200 Hz: there were no overlap of each pulse of plasma actuation as the frequency is less than 1.7 kHz as stated earlier. The size of SWBLI region has been defined as the distance between the bottom wall and the intersection of the shocks as shown in the figure. The initial difference between the experiment and calculation is about 0.2 mm which is within 0.22 mm of the experimental error[4]. In the parallel electrode case, the calculation and experiment indicated similar time profile: the reflected shock starts to move to downstream and the size of SWBLI region becomes large at around $t$ = 50 μs due to the heating of SWBLI region by inflow of the high temperature region as seen in Fig. 11. Then, the reflected shock goes back to the original position from $t$ = 250 or 300 μs. The heating effect appears to be overestimated in the calculation because the size of SWBLI region in the calculation is always larger than that in the experiment even considering the experimental error.

As for the canted electrode case, the heating effect or increase of the size of SWBLI region can be seen in the calculation from $t$ = 50 μs, while it cannot be observed in the experiment. The sizes begin to decrease in both calculation and experiment from around $t$ = 100 μs owing to the momentum transfer from the core flow to the boundary layer. The decrements in sizes are similar in both cases, about 0.5 mm. The size increases again immediately after that because of the heating effect in the calculation, however, the size remained small in the experiment. The heating effect seems to be overestimated in the calculation as well as the parallel electrode case, or the momentum transfer is underestimated.

Some discrepancies during plasma actuation between the experiment and calculation have been observed as discussed above. These are probably because of the assumption on the plasma region modeling; the size, shape and distribution of the temperature rise; however, the numerical calculation has demonstrated definite difference between the parallel and canted electrodes: the heating effect was dominant in the parallel electrode case, while the momentum transfer was enhanced in the canted electrode case. Therefore, qualitative investigations are available based on the calculation results since the calculation successfully captured the remarkable phenomena such as the heating effect and momentum transfer.

The evolution of the size of SWBLI region can be investigated in more detail from the velocity distribution. The $x$-velocity evolution is described in Fig. 13 in the parallel and canted electrodes cases. In both cases, the SWBLI region, or separation bubble becomes large as seen at $t$ = 78 μs because of the inflow of high temperature region into the SWBLI region. Then the bubble becomes larger in the parallel electrode case as seen at $t$ = 163 μs, while it becomes smaller than that in the initial in the canted electrode case as seen at $t$ = 258 μs due to the momentum transfer. Figure 14 (a) describes $x$-velocity profile in the boundary layer at $x$ = 0 mm (upstream of the SWBLI region, see Fig. 13) and $t$ = 98 μs. Flow acceleration in the boundary layer is definitely observed in the canted electrode case. The same plot at $x$ = 40 mm and $t$ = 248 μs is shown in Fig. 14 (b). The position of observation corresponds to the SWBLI region as depicted in Fig. 13. The heating effect and momentum transfer have been clearly confirmed in the plot comparing to the velocity profile in the case without plasma actuation: flow deceleration with a stronger backflow in the separation bubble can be observed in the parallel electrode case, indicating a stronger separation compared to the baseline without actuation, while there is very little backflow and even acceleration in the canted electrode compared to the baseline.

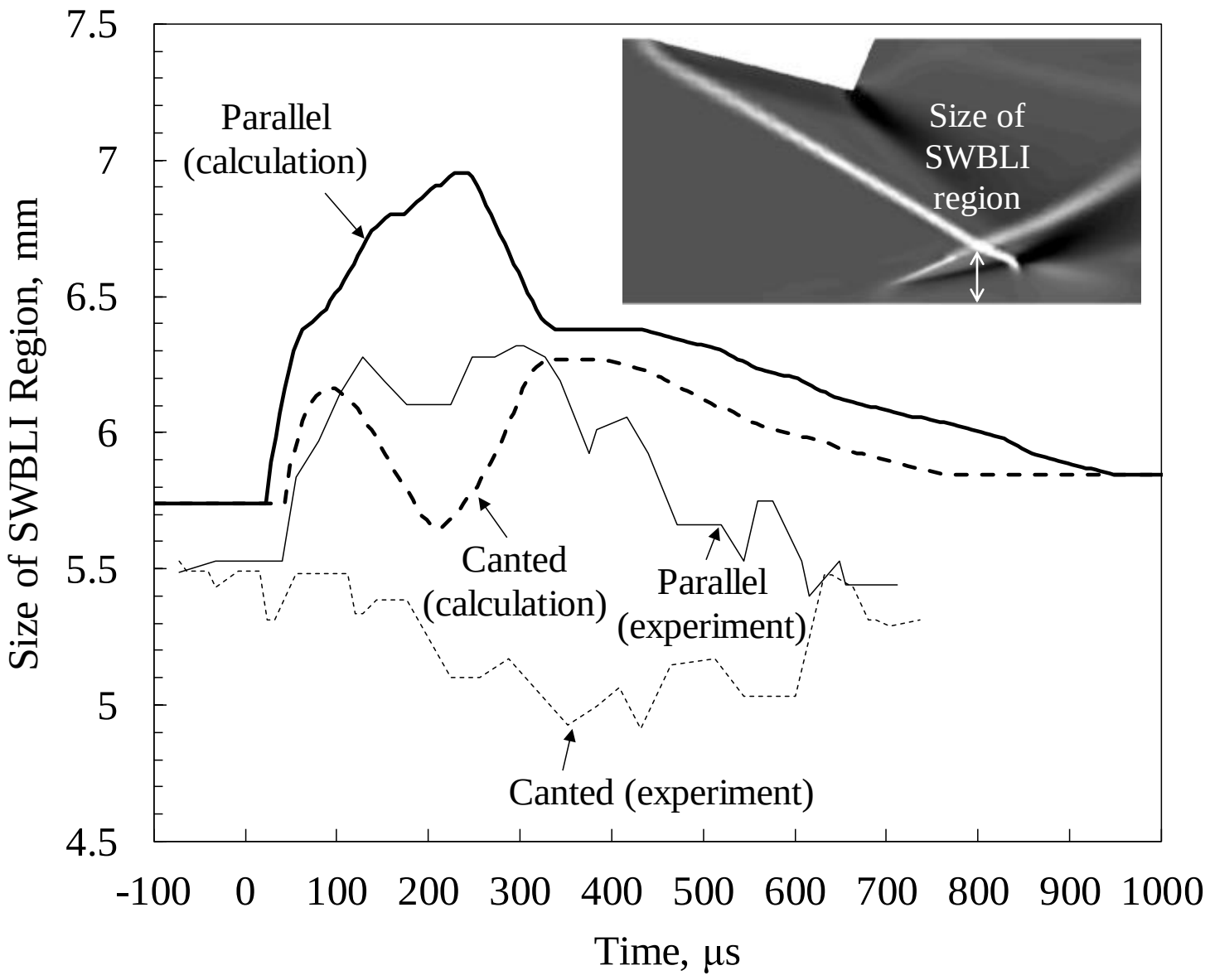

FIG. 12. Comparison of reflected shock movement as size of SWBLI region between calculation and experiment.

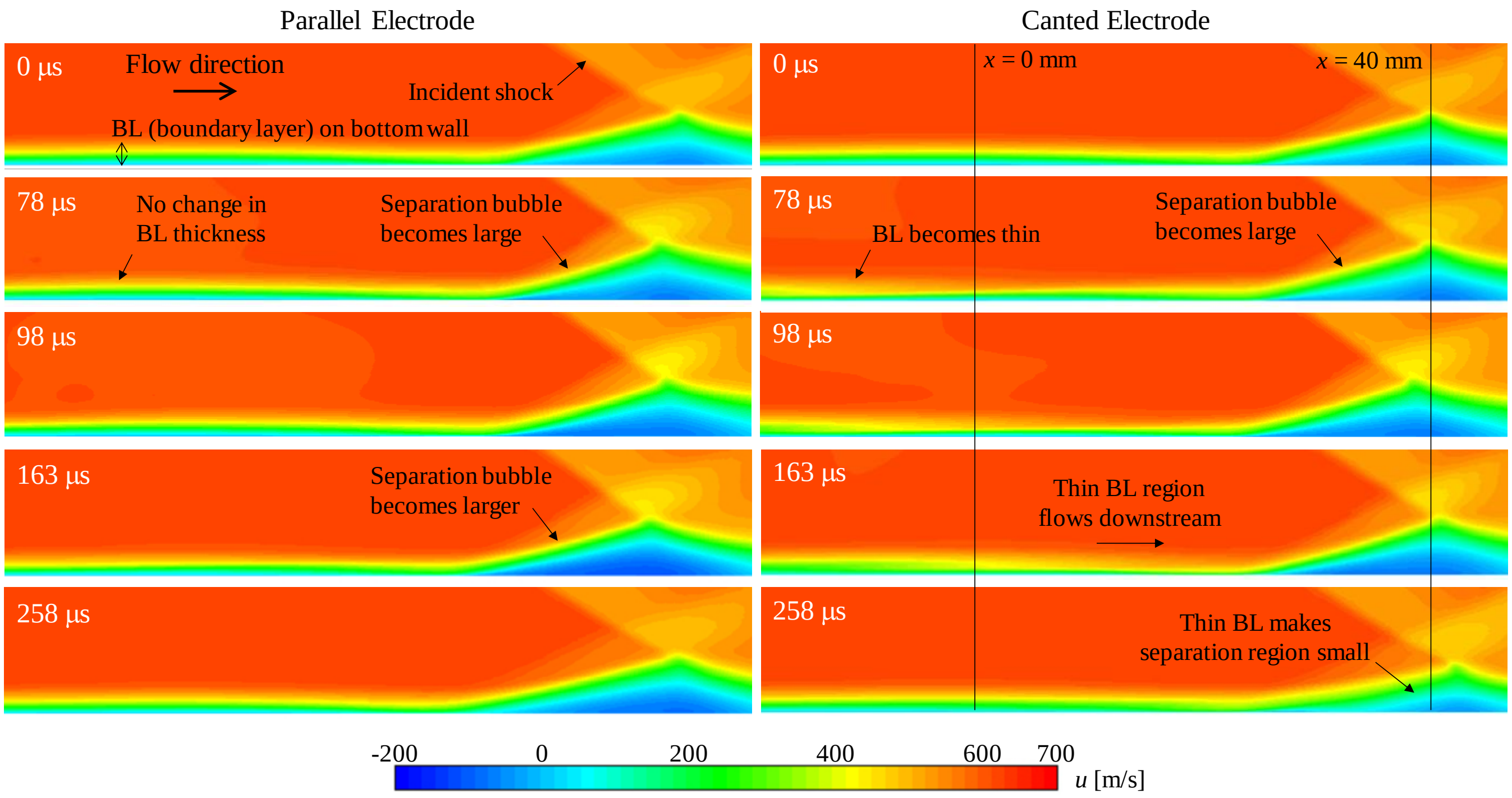

FIG. 13. *x*-direction velocity evolution at spanwise center in parallel and canted electrodes. Closed-up around and upstream of SWBLI region.

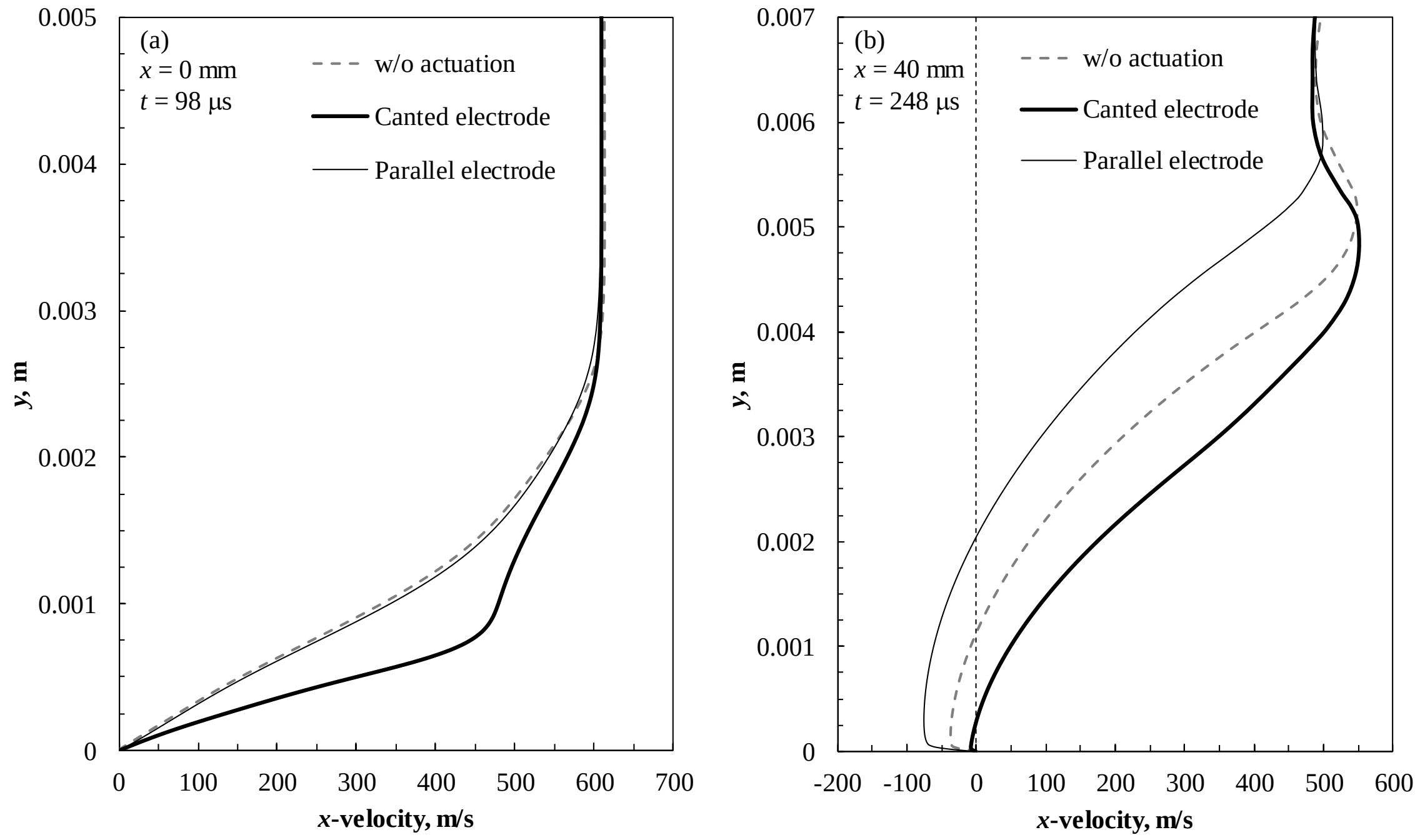

FIG. 14. *x*-direction velocity profile near bottom wall at spanwise center. (a) *x* = 0 mm, 98 μs and (b) at *x* = 40 mm 248 μs.

### *3. Mechanism of momentum transfer*

The mechanism of momentum transfer is investigated below. Figure 15 depicts *x*- and *z*-vorticities at $x$ = 0 mm cross section at 98 μs with *x*-velocity distribution and velocity vector in the parallel and canted electrodes cases. The vorticity vector is defined as follows.

$$\begin{pmatrix} \omega_x \\ \omega_y \\ \omega_z \end{pmatrix} = \begin{pmatrix} \frac{\partial w}{\partial y} - \frac{\partial v}{\partial z} \\ \frac{\partial u}{\partial z} - \frac{\partial w}{\partial x} \\ \frac{\partial v}{\partial x} - \frac{\partial u}{\partial y} \end{pmatrix} \quad (3)$$

In the parallel electrode case, there are no remarkable changes in the bottom boundary layer; however, in the canted electrode case, large vorticities in both *x*- and *z*-directions as well as a strong vortex in the bottom boundary layer occur. As seen in the velocity vector in the canted case, the strong vortex leads to momentum transfer into the boundary layer, consequently, the boundary layer becomes thin in the *x*-velocity contour. The strong vortex is generated by the interaction between main flow and induced flow by the actuator. Figure 16 describes *y*-direction velocity distribution at the spanwise center cross section as evidence of the momentum transfer process. Downward velocity near the bottom boundary layer that carries the momentum from the core flow can be frequently observed in the canted electrode case, on the other hand, there is no such velocity in the parallel electrode case.

The schematic summary of the momentum transfer mechanism is described in Fig. 17. In the canted configuration, the shock and expansion waves produced from the discharge region has the counter velocity against the main flow because they propagate perpendicular to the electrodes as depicted in Fig. 10. The counter flow plays an important role for vortex generation: the counter flow leads to *x*-velocity gradient in *y*-direction, or *z*-vorticity as indicated in Eq. (3). In the canted configuration, the generated *z*-vorticity interacts with the boundary layer as seen in the *z*-vorticity contour in Fig. 15. The *x*-vorticity has a similar distribution to the *z*-vorticity, which indicates the *x*-vorticity is induced through the interaction of *z*-vorticity. In other words, the discharge generates a thermal "bump" on the bottom wall; the flow around this bump interacts with the main flow and generates the vortex. The strong vortex induces the downward flow that conveys the core flow momentum to the boundary layer, as the result, a thinner boundary layer and smaller SWBLI region can be achieved as seen in both the experiment and calculation. In contrast, in the parallel electrode case, no counter flow has been seen as shown in Fig. 17; hence, no strong vortex has been observed in this configuration as shown in Fig. 15.

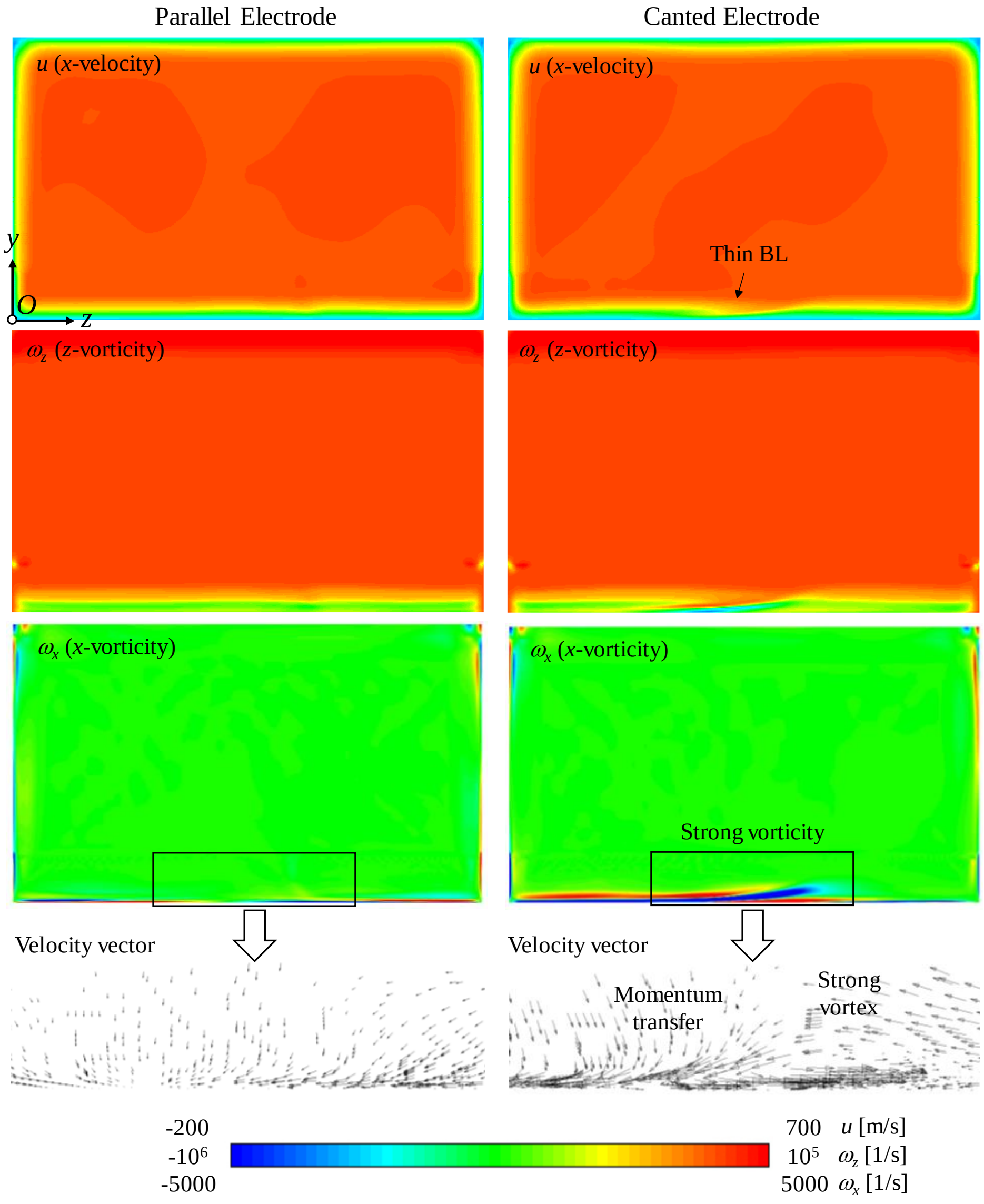

FIG. 15. Vorticities at $x$ = 0 mm cross section at 98 μs with $x$-velocity and velocity vector.

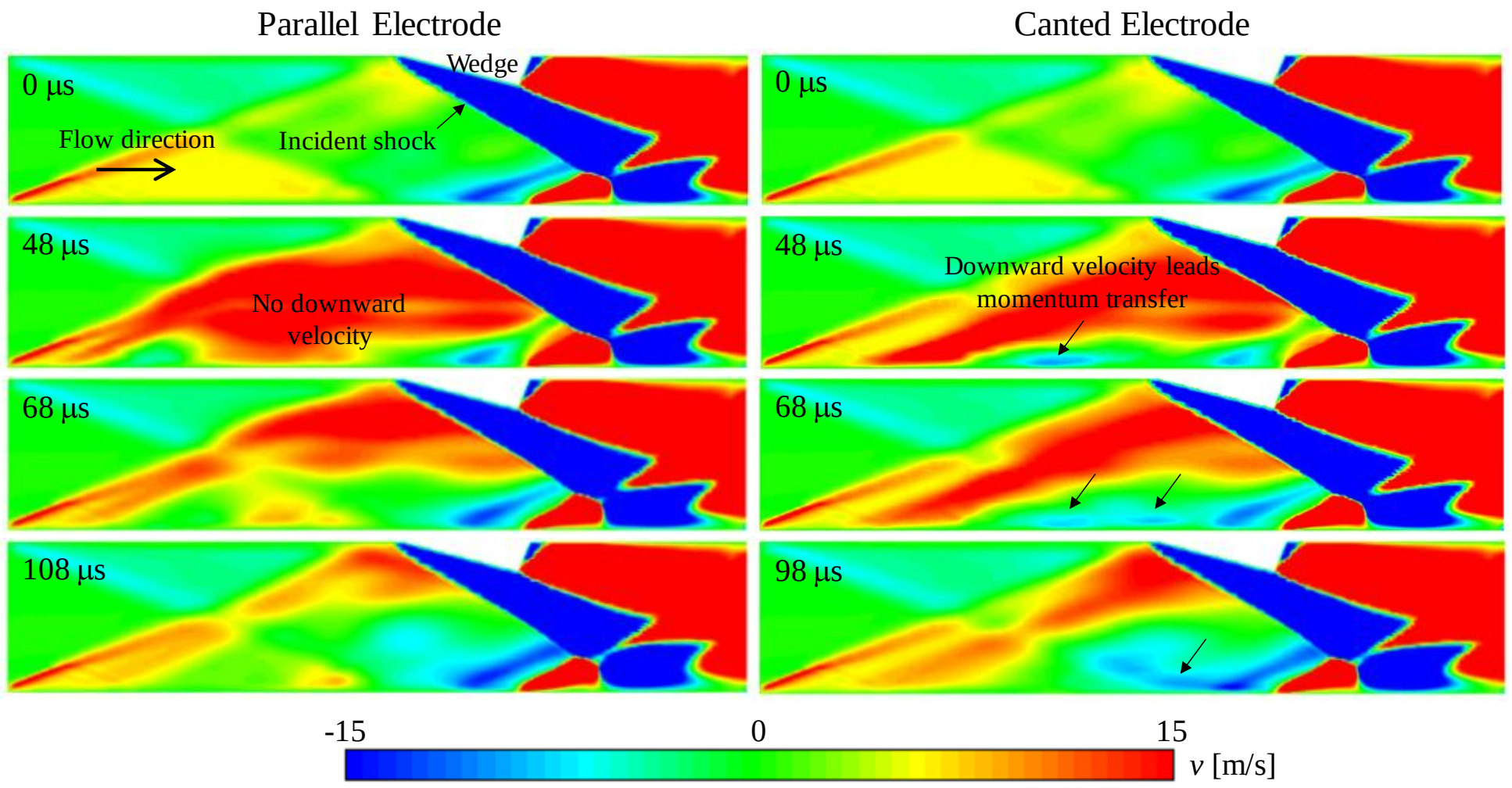

FIG. 16. *y*-direction velocity evolution at spanwise center in parallel and canted electrodes. Including whole calculation domain.

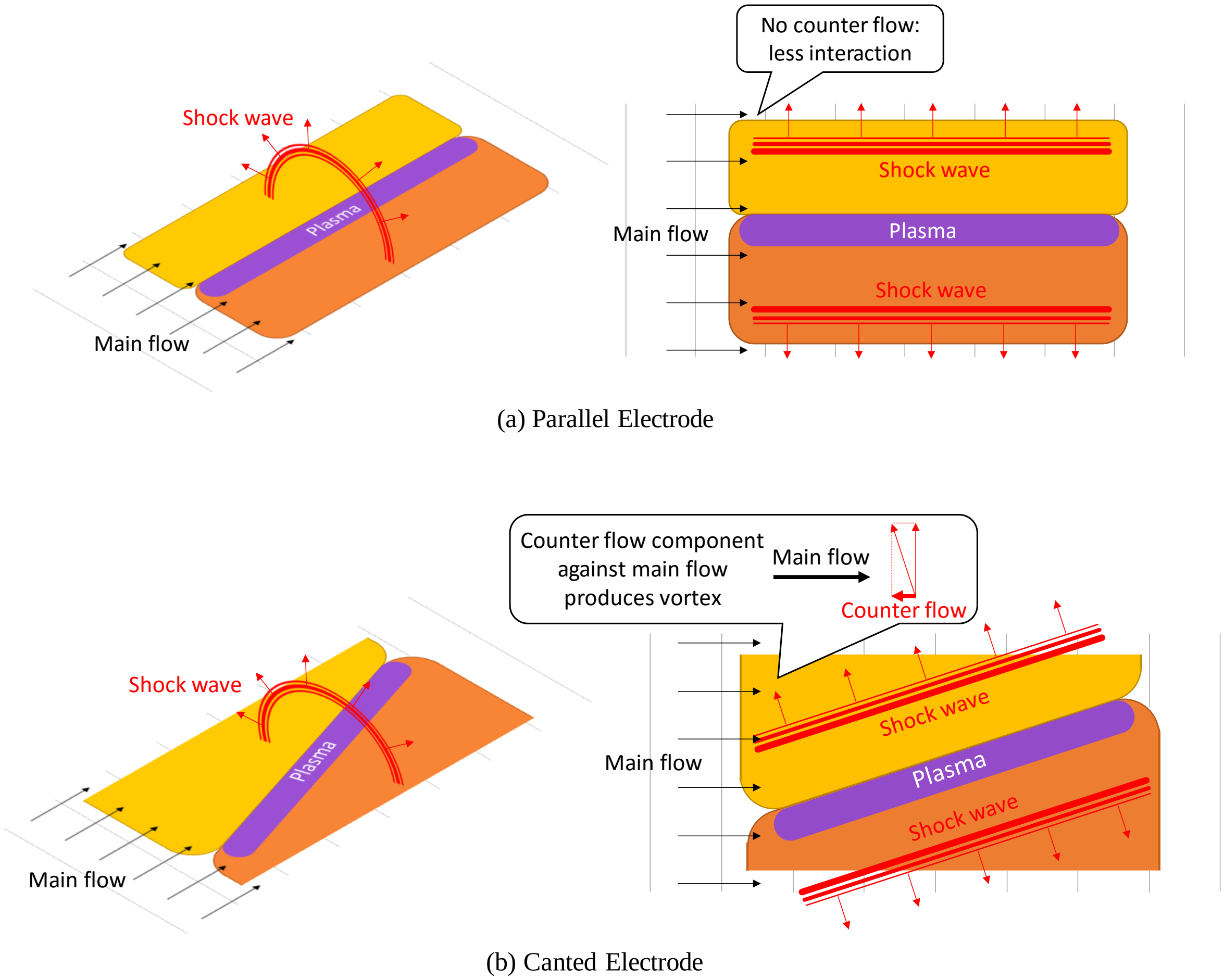

FIG. 17. Schematic of flow structure induced by parallel and canted electrodes.

## IV. CONCLUSION

To investigate the physical process in the NS-SDBD plasma actuator on the SWBLI, numerical calculations with a LES/RANS hybrid approach have been conducted. Firstly, the flowfield without the plasma actuation has been investigated and compared with the experimental data; wall pressure, schlieren image, and velocity vector. The comparison showed good agreement and the calculation result demonstrated the existence of a separation bubble and backflow.

An energy deposition model was adopted in which the hot plasma region was approximated as a 1/4 cylindrical high temperature region with a Gaussian temperature rise profile. The numerical investigation with the energy deposition model captured the remarkable features observed in the experiment: the heating effect and momentum transfer across the flow near the surface. As with the experiment, the parallel electrode causes a heating effect and a larger SWBLI region, while the canted electrode leads to a thinner boundary layer due to the momentum transfer. Discrepancies between the experiment and calculation has been observed in the size of SWBLI region probably due to the assumption in the hot plasma modeling. The velocity profiles in the boundary layer and the interaction region in the canted configuration demonstrated an acceleration in the boundary layer and a significant reduction of the backflow in the separation bubble. On the other hand, the flow deceleration and stronger backflow was observed in the parallel electrode case. The counter flow against the main flow produced from the high-temperature pulsed plasma plays an important role in the momentum transfer. Impingement between the main and counter flows makes a strong vortex near the boundary layer which convey the momentum to the boundary layer. The parallel electrode produces no counter flow, while the canted electrode does; thus, the thinner boundary layer and smaller SWBLI region has been observed only in the latter case.